\documentclass[useAMS,usenatbib]{mn2e}
\usepackage{graphicx}

\def\href#1#2{#2}
\def\urlinner#1{#1\endgroup}
\def\url{\begingroup\def\do##1{\catcode`##1 12 }%
  \do\\\do\$\do\&\do\#\do\^\do\_\do\%\do\~ \ttfamily \urlinner}

\usepackage[normalem]{ulem}

\def\apgt{\ {\raise-.5ex\hbox{$\buildrel>\over\sim$}}\,}
\def\aplt{\ {\raise-.5ex\hbox{$\buildrel<\over\sim$}}\,}

\title[A debris disk under the influence of a wide planet]{A debris disk under the influence of a wide planetary mass companion: The system of HD\,106906}

\author[Lucie J\'{i}lkov\'{a} and Simon Portegies Zwart]{Lucie J\'{i}lkov\'{a}\thanks{E-mail:
jilkova@strw.leidenuniv.nl (LJ); spz@strw.leidenuniv.nl (SPZ)} and Simon Portegies Zwart\footnotemark[1]\\
\noindent Leiden Observatory, Niels Bohrweg 2, Leiden, 2333\,CA, The Netherlands}
\begin{document}

\date{Accepted 2015 April 19. Received 2015 April 17; in original form 2014 July 02}

\pagerange{\pageref{firstpage}--\pageref{lastpage}} \pubyear{2002}

\maketitle

\label{firstpage}

\begin{abstract}
The 13\,Myr old star HD\,106906 is orbited by a debris disk of at least $0.067\,\rmn{M}_{\rmn{Moon}}$ with an inner and outer radius of 20\,AU and 120\,AU, respectively, and by a planet at a distance of 650\,AU. 
We use this curious combination of a close low-mass disk and a wide planet to motivate our simulations of this system. 
We study the parameter space of the initial conditions to quantify the mass loss from the debris disk and its lifetime under the influence of the planet.
We find that when the planet orbits closer to the star than about 50\,AU and with low inclination relative to the disk (less than about $10\degr$), more disk material is perturbed outside than inside the region constrained by observations on timescales shorter than 1\,Myr.
Considering the age of the system, such a short lifetime of the disk is incompatible with the timescale for planet--planet scattering which is one of the scenarios suggested to explain the wide separation of the planet.
For some configurations when the planet orbit is inclined with respect to the disk, the latter will start to wobble. 
We argue that this wobbling is caused by a mechanism similar to the Kozai--Lidov oscillations.
We also observe various resonant structures (such as rings and spiral arms) induced in the disk by the planet.
\end{abstract}


\section{Introduction}
\label{sec:intro}

About a dozen planetary mass companions at wide separations of about
50--100\,AU from their host stars have been revealed by direct imaging
surveys during the past decade \citep{kraus_three_2014} and several
cases were observed at separations of 150--300\,AU
\citep[e.g.,][]{lafreniere_direct_2008,kraus_three_2014}.  Moreover,
two recent discoveries report companions located as far as ${\sim}
650$\,AU \citep{bailey_hd_2014} and ${\sim} 2000$\,AU
\citep{naud_discovery_2014}. The origins of such wide planetary mass
companions is not well understood and presents important constraints
for our general understanding of planet formation. Several
scenarios have been proposed, and depending on the eccentricity and
separation of the planet, environment in which the system evolves, and
timescales of the formation, two main mechanisms are usually
considered.

In situ formation by core accretion
\citep[e.g.,][]{rafikov_constraint_2011} or protoplanetary disk
fragmentation \citep[e.g.,
][]{boss_formation_2011,vorobyov_formation_2013} can explain part of
the observed population of the wide orbit planets but is unlikely to
be the only formation channel (see also
\citealt{veras_formation_2009}, \citealt{dangelo_giant_2011}, or
\citealt{chabrier_giant_2014}, for recent reviews of the topic).

Another explanation argues that the planet formed closer to the
parent star in the protoplanetary disk and was scattered
outward by dynamical interaction with another planet
system or with perturbation of external origin (see e.g.,
\citealt{davies_long-term_2013}, for a summary on various interactions in
planetary systems). Given the diversity of the observed wide planetary
systems and the environment they are expected to form in, the
parameter space for the initial conditions of such scattering events
is extremely large and complex. The formation can involve for example,
stellar flybys \citep[e.g.,][]{malmberg_effects_2011}, exchange
interactions \citep{portegies_zwart_planets_2005}, planetary
migration \citep[e.g.,][]{crida_long_2009} and scattering in a multiple
planetary system \citep{scharf_long-period_2009}, dynamical
interaction between circumstellar disks and planets \citep[see][for a
  recent summary]{baruteau_planet-disc_2013}, the effects of Galactic
tides \citep[e.g.,][]{veras_exoplanets_2013}, recapture of free
floating planets \citep{perets_origin_2012}, or combination of these
interactions \citep{raymond_planet-planet_2010,boley_interactions_2012,hao_dynamical_2013}. Studying
specific objects narrows down this parameter space since some of the
characteristics are constrained by observations.

In this context, we focused on HD\,106906 which is a \mbox{F5-V} star with a debris disk \citep{chen_spitzer_2005,chen_magellan_2011} and a planetary mass companion at a distance of about 650\,AU \citep{bailey_hd_2014}. 
The chance of coincidental projection of the star and planet is negligible,
and therefore the observed distance between the star and the planet is interpreted as an orbital separation. Irrespective of the inclination of the planetary orbit, which is unknown, the observed separation must be part of the orbit, which makes it one of the widest separation ever observed.

Regardless of the process that caused this planet to have such a wide orbit, the observed debris disk has survived. 
The lifetime of the debris disk as is observed, constrains how long ago the current configuration formed.
In this paper we study the timescale on which the disk erodes due to the influence of the planet, and use this timescale to constrain the mechanism that delivered the planet in its extremely wide orbit.  
We carry out simulations of the evolution of the disk under the influence of the planet, taking the observed characteristics of the system as the initial conditions. 
We vary the inclination of the disk with respect to the planetary orbit and the pericenter distance of the planet (i.e., its eccentricity under the assumption that the apocenter distance of the orbit is 650\,AU) within the observational constraints, and we explore the erosion timescale of the disk due to the planet.

\subsection{The HD\,106906 system}
\label{sec:hd106906}

HD\,106906 (or also HIP\,59960) belongs to the Lower Centaurus Crux
(LCC) group which is a subgroup of the Scorpius--Centaurus (ScoCen) OB
association \citep{de_zeeuw_hipparcos_1999}.  The host star, called
HD\,106906\,A, is classified as \mbox{F5-V} star.
\citet{pecaut_revised_2012} measured the median age of the LCC group
of $17\pm 1\,\rmn{Myr}$, and the mass and age for HD\,106906\,A of
$M_{\star}=1.5\,\rmn{M}_{\sun}$ and $13\pm2\,\rmn{Myr}$, respectively.
In Table~\ref{tab:hd106906}, we summarize the observed data and
derived characteristics of the HD\,106906 system.

The observed infrared (IR) spectral energy distribution of
HD\,106906\,A shows a strong excess that indicates the presence of a
debris disk with inner cavity.  \citet[][see also
\citealt{chen_spitzer_2005}, for the initial results based on the
same observational data]{chen_magellan_2011} obtained broadband
observations of HD\,106906 with the Multiband Imaging
Photometer for Spitzer at 24 and 70$\,\umu$m.  By fitting these excess
fluxes with a single black-body, they derived the
disk's color temperature of 93$\,$K and fractional IR luminosity with
respect of the star $L_\rmn{IR}/L_{\star}=1.3 \times 10^{-3}$.
\citet{bailey_hd_2014} confirmed these results using additional
Spitzer data up to $\sim$100$\,\umu$m, obtaining a disk temperature of
95$\,$K.

The disk around HD\,106906\,A is expected to be optically thin.
\citet{chen_magellan_2011} identified 55 stars with IR excess in their
sample of 167 ScoCen OB Association members of intermediate-age
(10--30$\,$Myr) and F-, G-, or K- spectral types.  They did not find
any significant difference between the distribution of the IR excess
(measured by the $L_\rmn{IR}/L_{\star}$ ratio) for fast and slow
rotating stars. 
As a difference is expected in rotation speed for stars hosting gas-rich and gas-poor stars (due to magnetic braking, e.g., \citealt{rebull_correlation_2006}), it is likely that the stars in ScoCen association have optically thin and gas poor disks.

\begin{table}
  \caption{Characteristics of the HD~106906 system.}
  \label{tab:hd106906}
  \begin{center}
  \begin{tabular}{@{}lr@{$\pm$}llc@{}}
  \hline
  Characteristic & \multicolumn{2}{c}{Value} & Unit & Ref. \\
  \hline
  Distance & 92&2 &pc  & \textit{a} \\
  Age      & 13&2 &Myr & \textit{b} \\
  \hline
  \multicolumn{5}{c}{\emph{HD~106906~A}} \\
  Spectral type & \multicolumn{2}{c}{F5V} & & \textit{b} \\
  Mass $M_{\star}$            & 1.5&0.1  &$\rmn{M}_{\sun}$ & \textit{b} \\
  Luminosity $L_{\star}$      & 5.6&0.8  &$\rmn{L}_{\sun}$ & \textit{b} \\
  Temperature                 & 6516&165 &K                & \textit{b} \\
  \hline
  \multicolumn{5}{c}{\emph{HD~106906~b}} \\
  Mass $M_{\rmn{b}}$          & 11&2 &$\rmn{M}_{\rmn{Jup}}$ & \textit{c} \\
  Separation $R_{\rmn{b}}$    & 650&40&AU                   & \textit{c} \\
  \hline
  \multicolumn{5}{c}{\emph{disk}} \\
  24$\,\umu$m flux density                      & 103.1&2.5 & mJy                           & \textit{d} \\
  70$\,\umu$m flux density                      & 281&28    & mJy                           & \textit{d} \\
  Fractional luminosity $L_\rmn{IR}/L_{\star}$  & \multicolumn{2}{c}{$1.3\times 10^{-3}$} & & \textit{d} \\
  Dust grain temperature                        & \multicolumn{2}{c}{95}&K        & \textit{c} \\
  Inner radius                           & \multicolumn{2}{c}{$\sim$20} & AU         & \textit{c} \\
  Outer radius                           & \multicolumn{2}{c}{$<$120}   & AU          & \textit{c} \\
  Minimum mass                           & \multicolumn{2}{c}{0.067}    & $\rmn{M}_{\mathrm{Moon}}$  & \textit{d} \\
  \hline
  \end{tabular}
  \end{center}
  \medskip
  References: \textit{a}\,--\,\citet{van_leeuwen_validation_2007}, \textit{b}\,--\,\citet{pecaut_revised_2012}; \textit{c}~--~\citet{bailey_hd_2014}; \textit{d}\,--\,\citet{chen_magellan_2011}.
\end{table}

Since the disk is not resolved at any wavelength, its characteristic
extent can be estimated from the temperature. Assuming the dust grains
are black-bodies in radiative equilibrium with the central star,
an optically thin disk with grains of constant size and
chemical composition, \citet{chen_magellan_2011} derived a single
grain distance of about 34$\,$AU.  Based on the comparison with
Hershel observations of a sample of resolved circumstellar disks,
\citet{bailey_hd_2014} further estimated the extent of the disk to be about
20\,--\,120\,AU (for the optically thin disk).
\citet{chen_magellan_2011} also estimated the minimum dust grain size
of 1.4$\,\umu$m, and the minimum mass of the IR-emitting dust grains
of 0.067$\,\rmn{M}_{\rmn{Moon}}$.

The planetary mass companion of HD\,106906, called HD\,106906\,b, was
discovered by \citet{bailey_hd_2014} with the Magellan Adaptive
Optics/Clio2 system.  They obtained resolved images of the companion,
confirming that the planet is comoving with the host star, and
classified its spectral type as L2.5$\pm$1.  As mentioned above, 
the projected separation between the host star
and the companion then is 650\,$\rmn{AU}$.  Using evolutionary models
for an object of this spectral type and age corresponding to the one of the LCC
group, \citet{bailey_hd_2014} further estimated the mass of the planet
to be $M_{\rmn{b}}=11\pm$2\,$\rmn{M}_{\rmn{Jup}}$.
Properties of the planet make the formation of HD\,106906 difficult to explain. 
The two most compelling formation mechanisms for the origin of planets in wide orbits are discussed by \citet{bailey_hd_2014}: 
i)~in situ formation at a large separation, as wide as the orbital separation found in some binary stars; and 
ii)~formation in a tight orbit and the subsequent scattering to the current wide orbit. 
The mass ratio $M_{\rmn{b}}/M_{\star}\sim0.01$ is unusually small for the first suggested mechanism.
In the later scenario a perturber must have been present in order to move the planet to its current orbit. 
This culprit however, may be long gone, lost in interstellar space.  
This is consistent with the lack of another massive planet in the system \citep[][]{bailey_hd_2014}\,---\,no other object is detected within the observational limits which translate to a mass no grater $M_{\rmn{b}}$ beyond 35\,AU, and a mass no greater than 5--7\,$\rmn{M}_{\rmn{Jup}}$ beyond 70\,AU.
We cannot rule out other formation mechanisms, such as the possibly capturing of the planet from the surrounding environment in the LCC group.

Here we explore the lifetime of the current configuration of the system.
Planet--planet scattering is predicted to occur after the dissipation of the gas from the circumstellar disk at about $10^5$\,yr \citep[see e.g.][and references therein]{chatterjee_2008}.
Planets at wide separation ($>100$\,AU) are estimated to be most probably produced on timescales up to $10^7$\,yr \citep[e.g.,][]{veras_formation_2009,scharf_long-period_2009}.
If the current planetary orbit is the result of a scattering interaction with another planet, both planets once orbited the parent star in a much closer orbits, probably within the observed inner edge of the disk.
The current planetary orbit must still bear the memory of that original orbit and the place where the scattering happened, closer to the parent star, should also be part of the orbit.
The lifetime of the disk under the influence of such a planet should then be at least a few Myr in order to be consistent with the lifetime of the system.

We investigate the mass loss of the disk for different eccentricities and inclinations of the orbit with respect to the disk.


\section{Simulations}

We performed simulations of the evolution of the system starting with
initial conditions corresponding to its current observed
characteristics (see Table~\ref{tab:hd106906}).  We varied some of the
unconstrained properties, namely the pericenter of the planetary
orbit, $R_{\rmn{p}}$, and the inclination of the disk, $i$, since
these can in principle be random depending on the formation process of
the system. 

\subsection{Method}

We calculated the orbit of the star--planet system independently of
the evolution of the disk.  Since the mass of the disk is small compared
to the planet or the star, we represented the disk by a number of
zero-mass particles\,---\,planetesimals\,---\,and hence we do not take the self-gravity of the disk into account.

All calculations were
carried out within the Astrophysical Multipurpose Software Environment
or AMUSE
\citep{portegies_zwart_multiphysics_2009,pelupessy_astrophysical_2013}\footnote{\texttt{http://amusecode.org}}.
We used $N$-body integrator HUAYNO \citep{pelupessy_n-body_2012} to
calculate the orbit of the star--planet system.  The orbits of the
disk particles were calculated by solving the Kepler's equations using
universal variables \citep[adopted from the SAKURA
  code,][]{goncalves_ferrari_keplerian-based_2014}.  The
implementation of the solver in AMUSE allows us to efficiently
integrate Keplerian orbits in the potential of a central star with a
number orbiters (i.e., planetesimals orbiting the star in our case).
Our approach is not self consistent\,---\,the planet and the star are not influenced by the planetesimals in the disk.
The gravitational influence of the planet is coupled with the planetesimals.  
This coupling, called {\em Bridge} \citep{fujii_bridge:_2007}, is an
extension of the mixed variable symplectic scheme, which was developed
by \citet{wisdom_symplectic_1991}, and it is used here to couple
different dynamical regimes within one self-gravitating system (i.e.,
the planetesimal debris disk and the planet orbiting the central
star).  The time complexity of our numerical scheme is $\propto N$,
rather than the usual $\propto N^2$ for a direct \mbox{$N$-body}
approach.  The implementation of Bridge in AMUSE is described in
\citet{pelupessy_astrophysical_2013}.

The symplectic mapping method of \citet{wisdom_symplectic_1991} was
first applied to calculate the long-term evolution the solar system
and has since been widely used to simulate the evolution of planetary systems in
general, including interaction with planetesimals.  Fragmenting
planetesimals are generally considered to be the parent bodies of the
dust that is observed as a debris disk
\citep[e.g.,][]{wyatt_evolution_2008} and complex methods have been
developed to accurately model this process \citep[see, e.g.,][and
  references therein]{thebault_new_2012}.  The planetesimal disk
approximation is often used to define the spatial and velocity
distributions of the dust particles.  For example,
\citet[][]{larwood_close_2001} investigated the affect of stellar
flybys on the structure of the debris disk observed in the
$\beta$\,Pictoris system, and similarly in
\citet{chiang_fomalhauts_2009} for the Fomalhaut system.
\citet{wyatt_resonant_2003} or \citet{reche_observability_2008}
studied the resonant trapping of planetesimals due to planetary
migration.  \citet{lestrade_stripping_2011} investigated the stripping
of the planetesimal debris disk by a close stellar flyby. 
Long-lived asymmetric structures were simulated by, e.g., \citet[][eccentric debris disk around $\zeta^2$\,Reticuli]{faramaz_can_2013} or \citet[][more general case of a planet within the outer edge of the disk]{pearce_dynamical_2014}.

We tested the method by comparing our implementation with direct $N$-body integrations, which gave qualitatively and quantitatively the same results; and we successfully reproduced the results of \citet{lestrade_stripping_2011}.

\subsection{Numerical setup and initial conditions}
\label{sec:num_setup}

Following the observations, we assumed a mass of
$1.5\,\rmn{M}_{\sun}$ for the star and $11\,\mathrm{M}_{\rmn{Jup}}$
for the planet (see Sect.~\ref{sec:hd106906} and
Table~\ref{tab:hd106906}).  The apocenter distance of the planet was
650\,AU in all our simulations. 
This is the observed separation, which we assume to be the apocenter of the orbit, and which is the planet's initial position in our simulations.
The pericenter distance of the planet, $R_{\rmn{p}}$,
had values ranging from 1$\,$AU to 650$\,$AU, corresponding to orbital
eccentricities of 0.997 and to circular orbit, respectively (see
Table~\ref{tab:bin_orbit} for the list of all pericenter values
considered).  The orbit of the planet was integrated with HUAYNO using
the HOLD drift--kick--drift integrator.  The HUAYNO integrator uses
individual time-steps that are proportional to inter-particle
free-fall times and the coefficient of the proportionality is called
$\eta$.  We chose different values of $\eta$ for different pericenters
(i.e., orbital eccentricities) so that the energy conservation of the
star--planet system is always at $10^{-6}$ level and lower; this level
of energy conservation turns out to be very conservative
\citep{portegies_zwart_minimal_2014}. The values of $\eta$ are
specified in Table~\ref{tab:bin_orbit} for each orbital configuration.

\begin{table}
  \begin{minipage}{0.45\textwidth}
  \caption{Planetary pericenters and time-steps for the integrations.}
  \label{tab:bin_orbit}
  \centering
  \begin{tabular}{ccc}
    \hline
    $R_{\rmn{p}}$ [AU] & $\eta$ & $t_{\rmn{BR}}$ \footnote{The Bridge time-step, $t_{\rmn{BR}}$, is given in the units of the period of the circular orbit at $20\,\rmn{AU}$ from the star, which is $73\,\rmn{yr}$.}\\
    \hline
    1                            & 0.001 & 0.001 \\
    10                           & 0.001 & 0.002 \\
    20, 30, 40, 50, 60           & 0.001 & 0.01 \\
    70, 80, 90, 100, 110         & 0.001 & 0.05 \\
    120, 150, 200, 350, 500, 650 & 0.003 & 0.05 \\
    \hline
  \end{tabular}
  \vspace*{-0.5cm}
  \end{minipage}
\end{table}

Disk planetesimals begin in an initially a uniform random
distribution in radius between the inner and outer disk radii of
20$\,$AU and 120$\,$AU, respectively, which corresponds to the values
estimated from observations (see Sect.~\ref{sec:hd106906} and
Table~\ref{tab:hd106906}). 
Such choice of radial distribution corresponds to the surface density profile $\propto 1/r$, where $r$ is the radial distance to the star, which is often used to model proto-planetary disks \citep[see e.g.,][and references therein]{steinhausen_2012} and is supported by observations \citep[e.g.,][]{andrews_2007}.
Following the discussion in \citet{steinhausen_2012}, we tested how our results depend on the chosen initial surface density profile. Since the disk is represented by test particles (i.e., its self-gravity is not taken into account), different surface density profiles can be taken into account in the post-processing of the simulations. We considered a surface density profile $\propto 1/r^{1.5}$, corresponding to the Minimum Mass Solar Nebula \citep{hayashi_1981}, and we found that such profile changes the disk fractions presented in Sec.~\ref{sec:parameter_space} by less than 10\%.

The planetesimals are initially placed in one plane with a random, uniform azimuthal distribution and they have circular orbits. The inclination of the disk with respect to the planetary orbit, $i$, has values between 0$\degr$ and 180$\degr$, where $i=0\degr$ corresponds to coplanar prograde case, 
and $i=180\degr$ corresponds to coplanar retrograde case. 
The disk plane is rotated
around axis perpendicular to semi-major axis of the planetary orbit.
Each simulation was carried out with $10^4$ particles, but we
confirmed that increasing this number to $10^5$ does not change the
results.  Decreasing the number to $10^3$ particles gives
qualitatively similar results, but the smaller number of particles
makes post processing analysis harder due to the lower statistics.

The planetesimals feel the gravitational force from the planet with
specific time-step of the interaction, $t_{\rmn{BR}}$\,---\,the Bridge
time-step\,---\, which is the time step in which the system integrates
the combined solver. The time-step differs for different initial
conditions of the planetary orbit\,---\,for more eccentric
planetary orbits we adopted a shorter time-step. $t_{\rmn{BR}}$ has
values ranging from $10^{-3}$ to $5\times10^{-2}$ of an orbital period
of the initial inner disk edge of 73\,$\rmn{yr}$ (which is the case
for the adopted 20\,$\rmn{AU}$).  The values of $t_{\rmn{BR}}$ are
specified in Table~\ref{tab:bin_orbit} for each orbital configuration.
We verified the choice of $t_{\rmn{BR}}$ by comparing the integrations
using Bridge to the calculations where the whole system was treated by the
$N$-body code.  These control $N$-body simulations were carried with
$10^3$ zero-mass particles in the disk.  We used the HUYANO integrator
in AMUSE with choice of $\eta$ giving the energy conservation of order
$10^{-6}$ or lower.  To treat close encounters of the planetesimals
with the star, we use Plummer softening with smoothing length
$\epsilon = 0.001\,\rmn{AU} = 0.2\,\rmn{R}_{\sun}$. 
The results of the direct and our Bridged direct--Kepler solver are in a good agreement. 
More quantitatively, we compared the disk fraction, $f_{\mathrm{d}/\mathrm{b}}$\,---\,the main output of our simulations defined in Sect.~\ref{sec:parameter_space}\,---\,which generally agrees on a ${\sim} 5$\% level.

\section{Results}

\begin{figure*}
  \includegraphics[width=\textwidth]{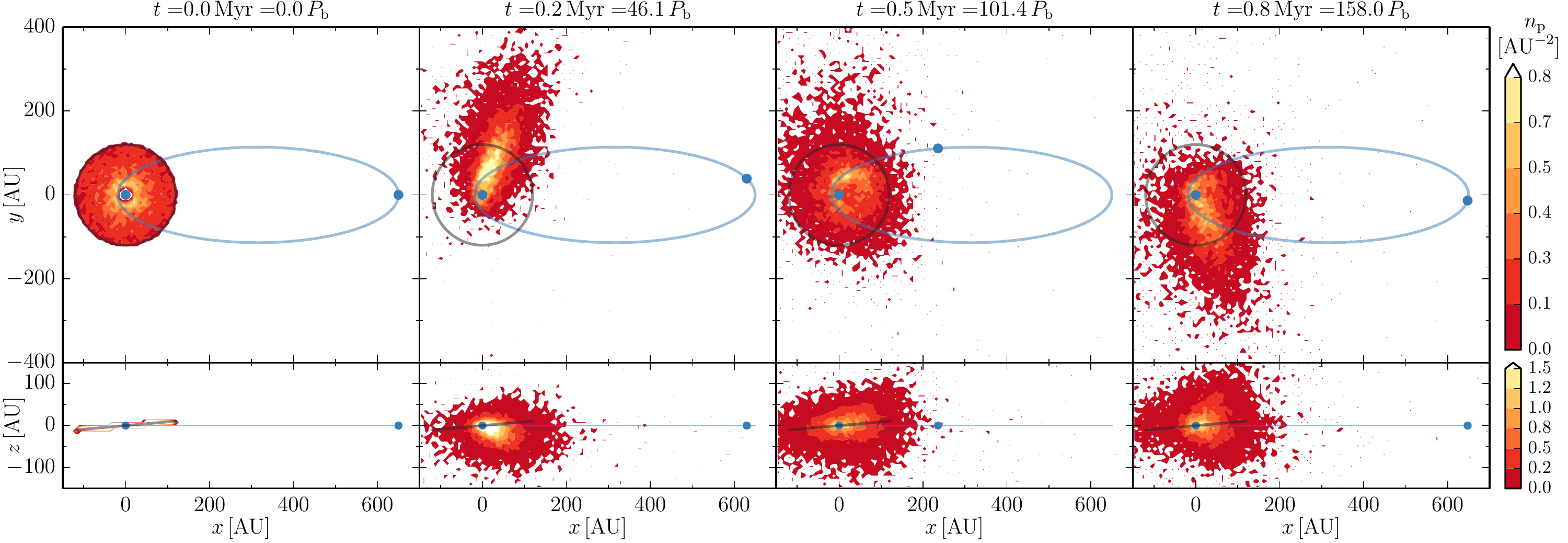}
  \caption{Snapshots from the simulation with $R_{\rmn{p}}=20$\,AU and $i=5\degr$. 
  Time of the snapshots is indicated above each panel in Myr and in $P_{\rmn{b}}$ (orbital period of the planet). 
  The color-scale maps the number of planetesimals, $n_{\rmn{p}}$, projected in the planetary orbit plane ($xy$) and the edge-on view of the initial disk (plane perpendicular to the planetary orbit, $xz$) in the upper and the lower panel respectively. 
  The star, the planet and its orbit are indicated in blue. 
  The gray ellipse and line segment show the initial extend of the disk. 
  The planet and the planetesimals rotate in the same sense, counter clock-wise in the $xy$ plane.}
  \label{fig:snaps}
\end{figure*}

In Fig.~\ref{fig:snaps}, we show an example of our simulation\,---\,the configuration with pericenter of 20\,AU (at the inner edge of the disk) and the disk inclination of $5\degr$. 
The surface number density of planetesimals in the plane of the planetary orbit ($xy$) and the edge-on plane ($xz$) is plotted in the upper and lower panel respectively.
As the planet plunges through the disk, it perturbs the planetesimals' orbits and the disk is disrupted.
Some planetesimals move outside the initial disk region and some become unbound from the star and escape from the system.
The majority of the particles that are moving outside the initial disk region are perturbed farther away from the star, i.e. their semi-major axis is larger than the outer disk radius of 120\,AU (indicated by the gray ellipse in Fig.~\ref{fig:snaps}), and only a small fraction of particles are orbiting within the inner disk edge (with semi-major axis smaller then 20\,AU). 
Note that we do not consider collisions between the planetesimals themselves neither with the star nor the planet and no particles are removed from the simulation.

\subsection{Parameter space study}
\label{sec:parameter_space}

\begin{figure*}
  \includegraphics[width=0.65\textwidth]{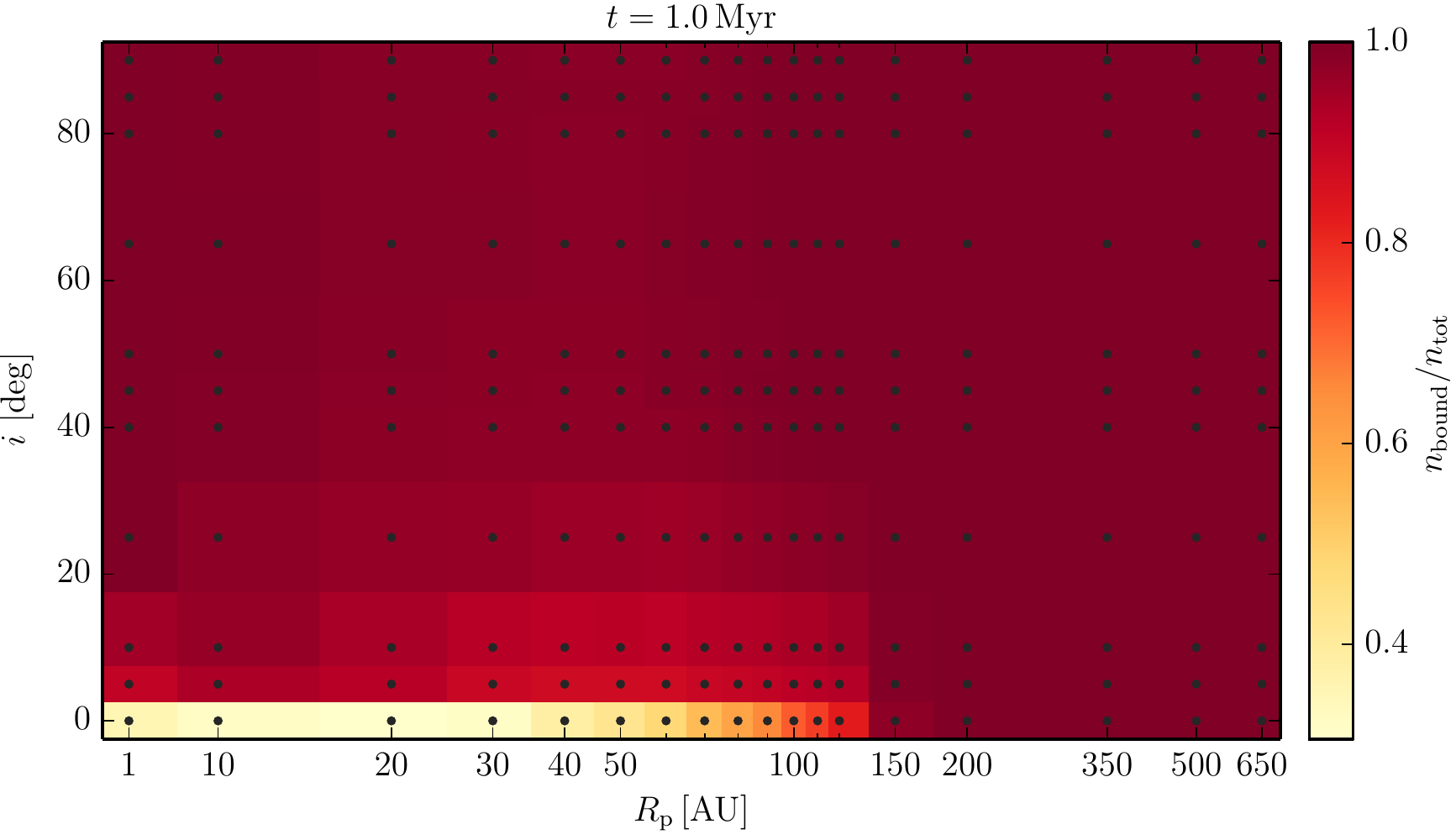}
  \caption{The fraction of particles that stay bound to the star after 1\,Myr mapped in the pericenter--inclination plane. 
  The planetary pericenters and disk inclinations are changing along the horizontal and the vertical axis, respectively.
  The plane is divided in colored bins and the $R_{\rmn{p}}$ and $i$ of the used grid are indicated by points.
  Note that the horizontal axis is logarithmic except for the smallest pericenter (1\,AU), which is shown in different scale for clarity.}
  \label{fig:nb_ntot}
\end{figure*}

We explored the parameter space of the pericenter of the planetary orbit ($R_{\rmn{p}}$) and the inclination of the disk with respect to the orbital plane ($i$).
In Fig.~\ref{fig:nb_ntot} we show the fraction of the disk particles that stay bound to the star after 1\,Myr of the evolution\,---\,$n_{\rmn{bound}}/n_{\rmn{tot}}$, where $n_{\rmn{bound}}$ is the number of bound particles and $n_{\rmn{tot}}$  is the total number (i.e., $n_{\rmn{tot}}=10^4$).
Fig.~\ref{fig:nb_ntot} maps the prograde cases ($0\degr<i\leq90\degr$); the results for the retrograde configurations are generally similar (see below for some examples).
We see that only in the coplanar case when the pericenter is smaller that the outer disk radius, a substantial number of particles is lost (unbound) from the system.
It is hardly surprising that the highest number of unbound particles is produced in such configurations, 
but it is interesting that more than $\sim80$\% of the particles stays bound for all the other considered configurations during the first 1\,Myr.

\begin{figure*}
  \includegraphics[width=0.4333\textwidth]{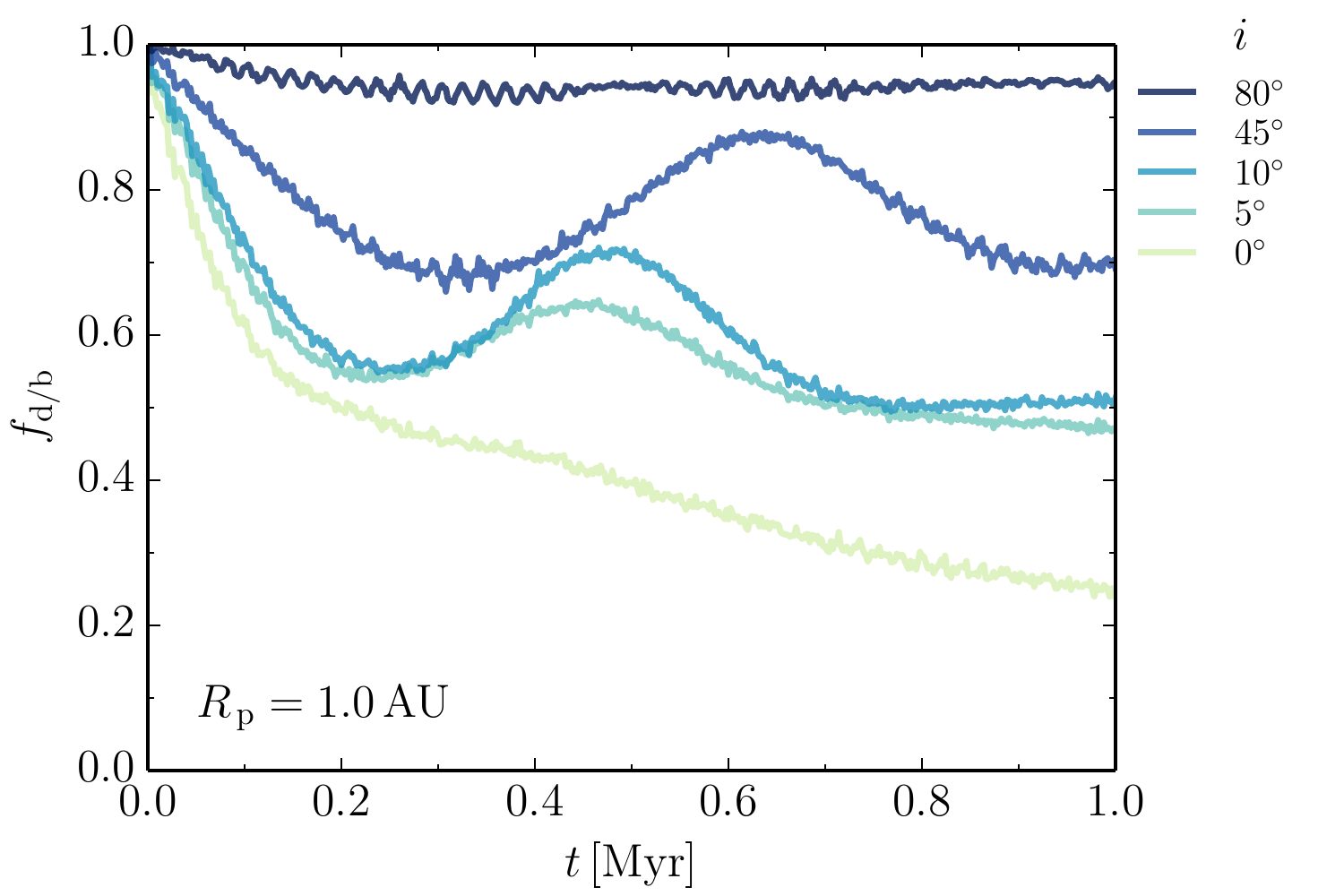}
  \includegraphics[width=0.4333\textwidth]{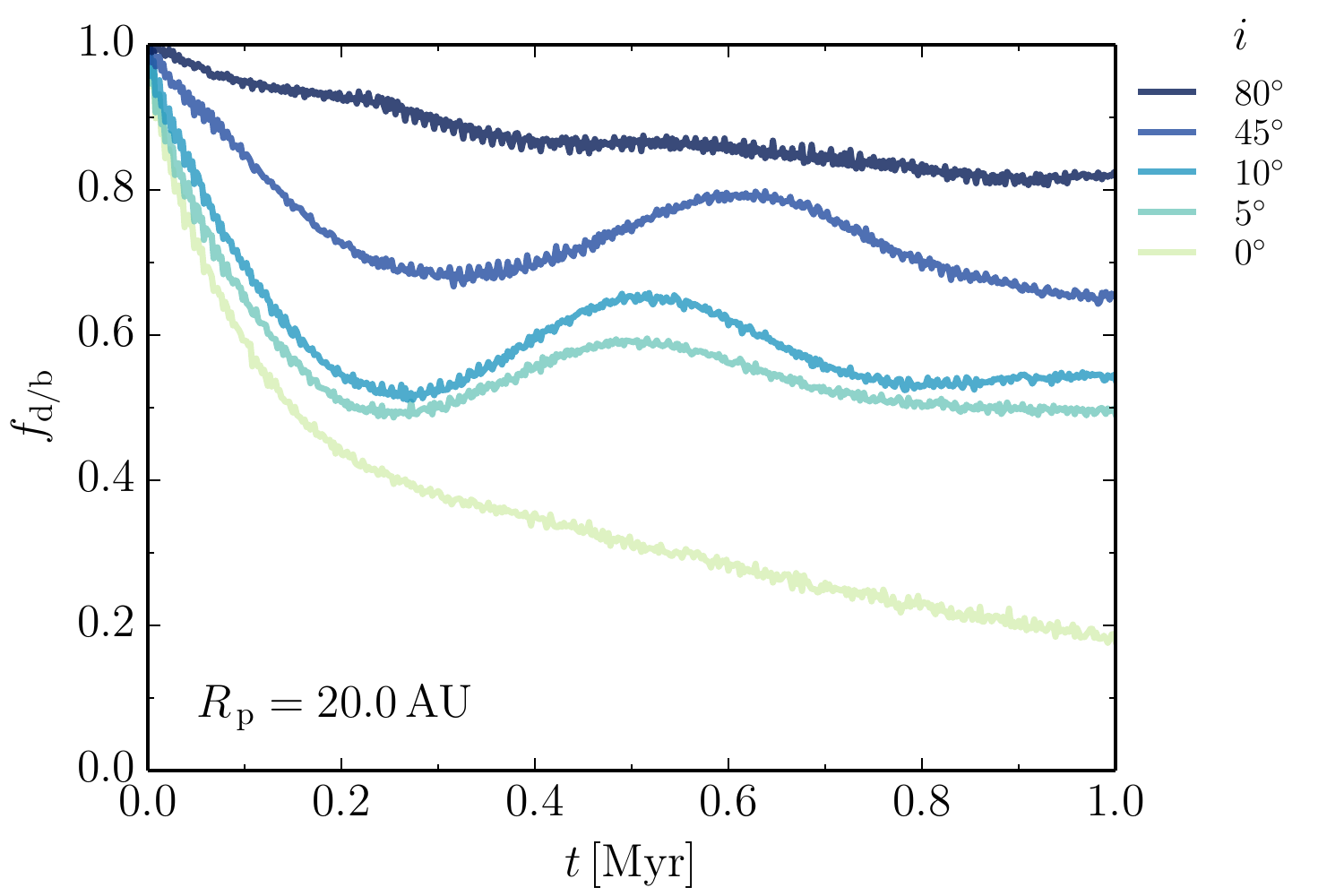} \\
  \caption{Evolution of the fraction $f_{\mathrm{d}/\mathrm{b}}(t)$ for a pericenter distance of 1\,AU (left) and 20\,AU (right) and various inclinations of the disk with respect to the planetary orbit $i<90\degr$ (prograde cases). 
  The lines of different colors correspond to different $i$ as indicated to the right of each plot.}
  \label{fig:ndot_varchar}
\end{figure*}

The number of bound particles measures what part of the original disk is kept within the system which, however, does not directly correspond to the observed disk.
For example, in the second and the last snapshots of Fig.~\ref{fig:snaps}, we see that a substantial number of the planetesimals is located outside the disk area as it was constrained from the observations.
Most of these planetesimals are however still bound to the star and the ratio $n_{\rmn{bound}}/n_{\rmn{tot}}$ is about 0.8 at 1\,Myr (see Fig.~\ref{fig:nb_ntot}).
Majority of these bound particles perturbed from the disk extent have semi-major axis larger than the outer edge of the disk of 120\,AU, while only small fractions orbits within the inner edge.

To estimate how consistent our simulations are with the observed disk, we follow the ratio of the number of particles with their instantaneous distance from the star within the observationally constrained disk extent and the number of particles bound to the star.
We call this quantity {\it disk fraction} $f_{\mathrm{d}/\mathrm{b}}$ and it is given as $f_{\mathrm{d}/\mathrm{b}} = n(20\,\rmn{AU}<R<120\,\rmn{AU}) / n_{\mathrm{bound}}$, where $n(R)$ is the number of particles at given distance $R$ (spherical radius) from the star.
We use the instantaneous distance because the disk is not resolved in the observations and its extent is estimated from the temperature that is given by the distance of the debris from the star.
We tested that in case when the semi-major axis of the particles' orbits is used instead of the instantaneous distance, the evolution of the ratio stays generally similar however, its modulations, both the short- and the long-term (see Sec.~\ref{sec:discussion}), are not present.

As mentioned, the ratio $f_{\mathrm{d}/\mathrm{b}}$ measures the similarity of the simulated system to the observed state.
If this ratio is high, most of the particles are orbiting within the radii constrained by observations;
low value of $f_{\mathrm{d}/\mathrm{b}}$ indicates that most of the particles bound to the star are orbiting outside the constrained radii. 

In Fig.~\ref{fig:ndot_varchar}, we show the evolution of $f_{\mathrm{d}/\mathrm{b}}$ over 1\,$\rmn{Myr}$ for the cases when the pericenter of the planetary orbit is $1\,\rmn{AU}$ and when it coincides with the inner edge of the disk ($R_{\rmn{p}}=20\,\rmn{AU}$) for a number of disk inclinations.
We focus on the cases with the pericenter within the inner disk edge because such configurations 
are expected if the planetary orbit is the result of a planet--planet scattering.
In both cases, generally the lower the inclination, the lower the ratio $f_{\mathrm{d}/\mathrm{b}}$ and there is about 30\% difference between the inclination of $5\degr$ and the coplanar configuration.
The evolution of $f_{\mathrm{d}/\mathrm{b}}(t)$ is not monotonic and is subject of (at least) two modulations with different timescales of about 0.05 and 0.3\,Myr.

\begin{figure}
  \includegraphics[width=0.4333\textwidth]{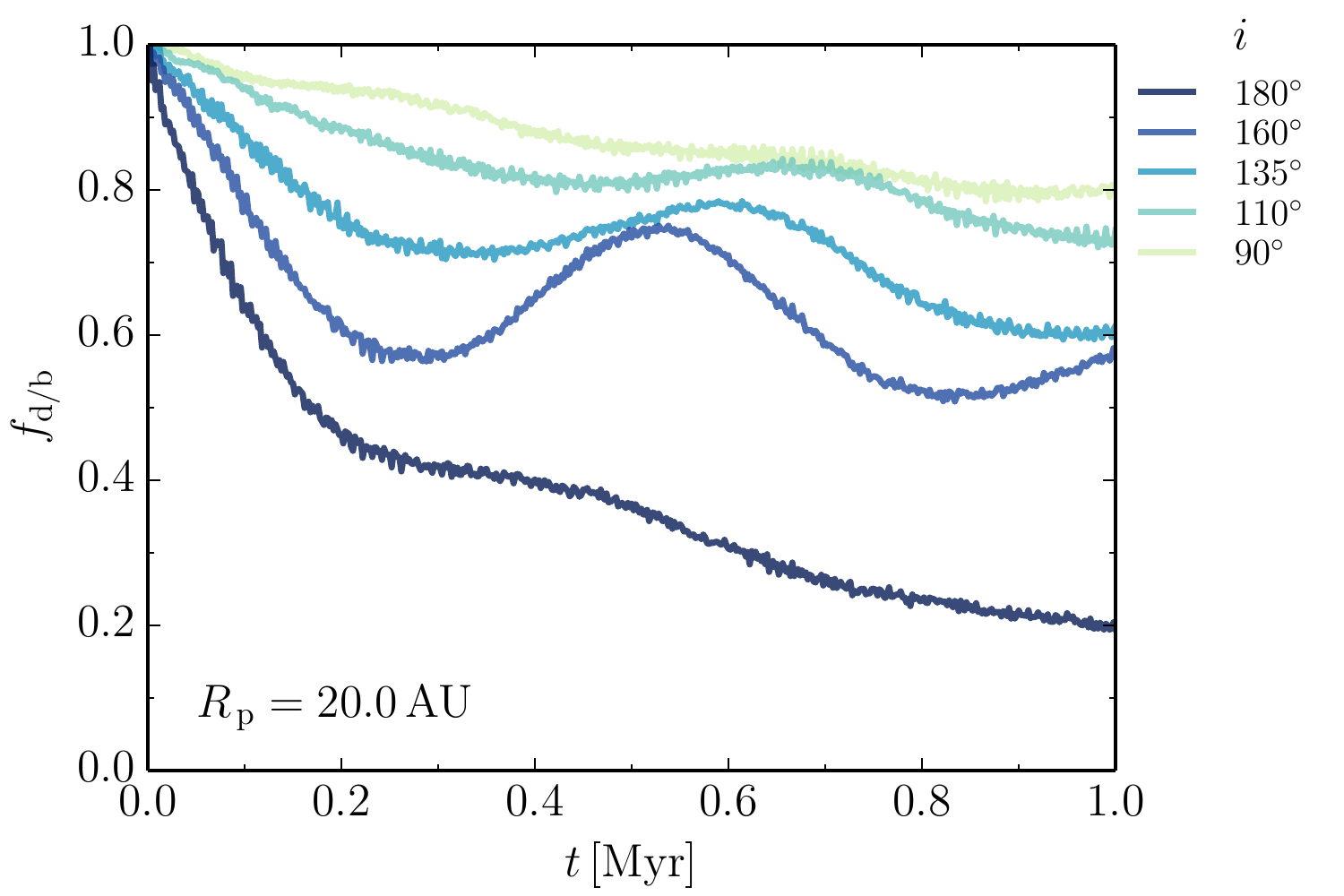}
  \caption{Evolution of the fraction $f_{\mathrm{d}/\mathrm{b}}(t)$ for a pericenter distance of 20\,AU and various inclinations of the disk with respect to the planetary orbit $i\geq90\degr$ (retrograde cases).
  }
  \label{fig:ndot_retro}
\end{figure}

In Fig.~\ref{fig:ndot_retro}, we show $f_{\mathrm{d}/\mathrm{b}}(t)$ for configurations when the disk has a retrograde rotation with respect to the orbit of the planet (i.e., $i\geq90\degr$) with pericenter of 20\,AU.
The evolution of the disk fraction looks generally very similar to the prograde cases with the same planetary pericenter (Fig.~\ref{fig:ndot_varchar}, right).

\begin{figure*}
  \includegraphics[width=0.4333\textwidth]{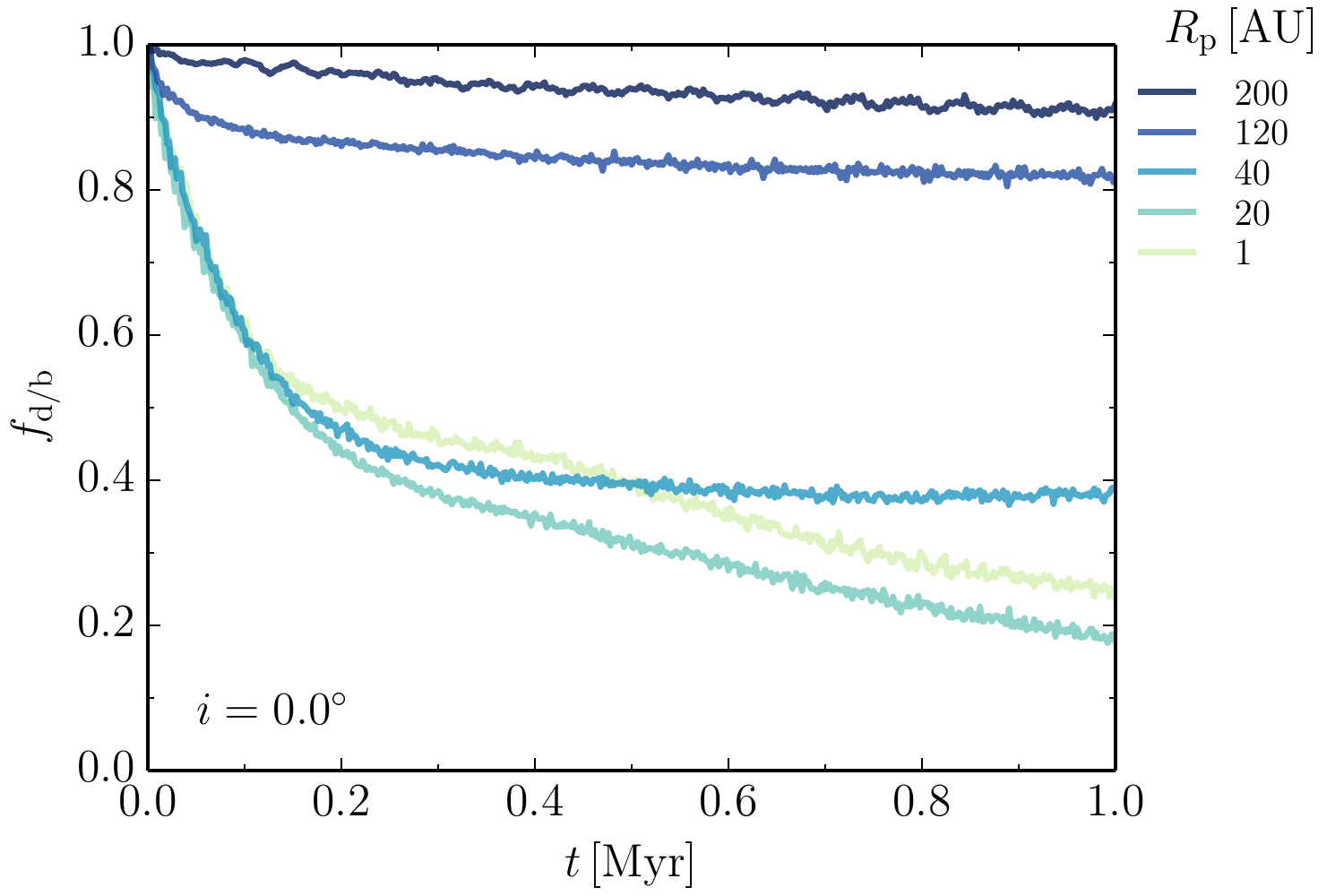}
  \includegraphics[width=0.4333\textwidth]{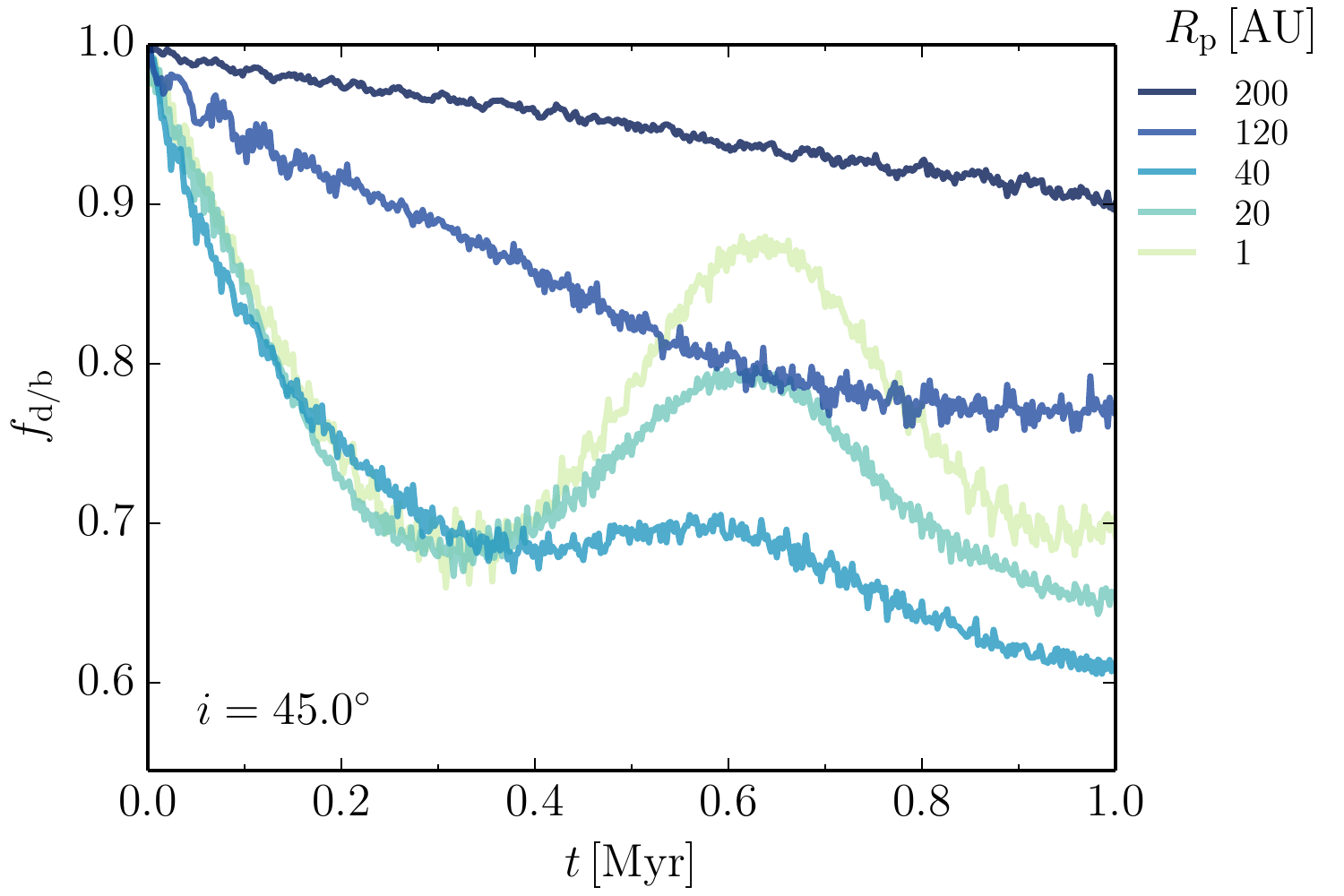} \\
  \caption{Evolution of the fraction $f_{\mathrm{d}/\mathrm{b}}(t)$ for disk's inclination of $0\degr$ (left) and $45\degr$ (right) and various pericenters of the planetary orbit. 
  The lines of different colors correspond to different $R_{\rmn{p}}$ as indicated to the right of each plot.
  }
  \label{fig:ndot_fixed_i}
\end{figure*}

Finally, in Fig.~\ref{fig:ndot_fixed_i}, we show $f_{\mathrm{d}/\mathrm{b}}(t)$ for fixed inclinations of $0\degr$ and $45\degr$ and several values of the pericenter of the planetary orbit.
As expected, the disk fraction is generally higher for the configurations with larger pericenters\,---\,more than about 80\% of the particles is within the disk for paricenters beyond the outer edge, $R_{\rmn{p}}>120$\,AU.
Similarly as in Figs.~\ref{fig:ndot_varchar} and \ref{fig:ndot_retro}, the disk fraction oscillates with two different timescales\,---\,the modulation with the longer timescale occurs only in cases with non-zero inclination, while the shorter one is present for configurations with higher disk fraction $f_{\mathrm{d}/\mathrm{b}}\apgt 0.7$.
The possible explanation of these is discussed in Sec.~\ref{sec:discussion}.

\subsection{Disk lifetime}
\label{sec:disk_lifetime}

\begin{figure}
  \includegraphics[width=0.4333\textwidth]{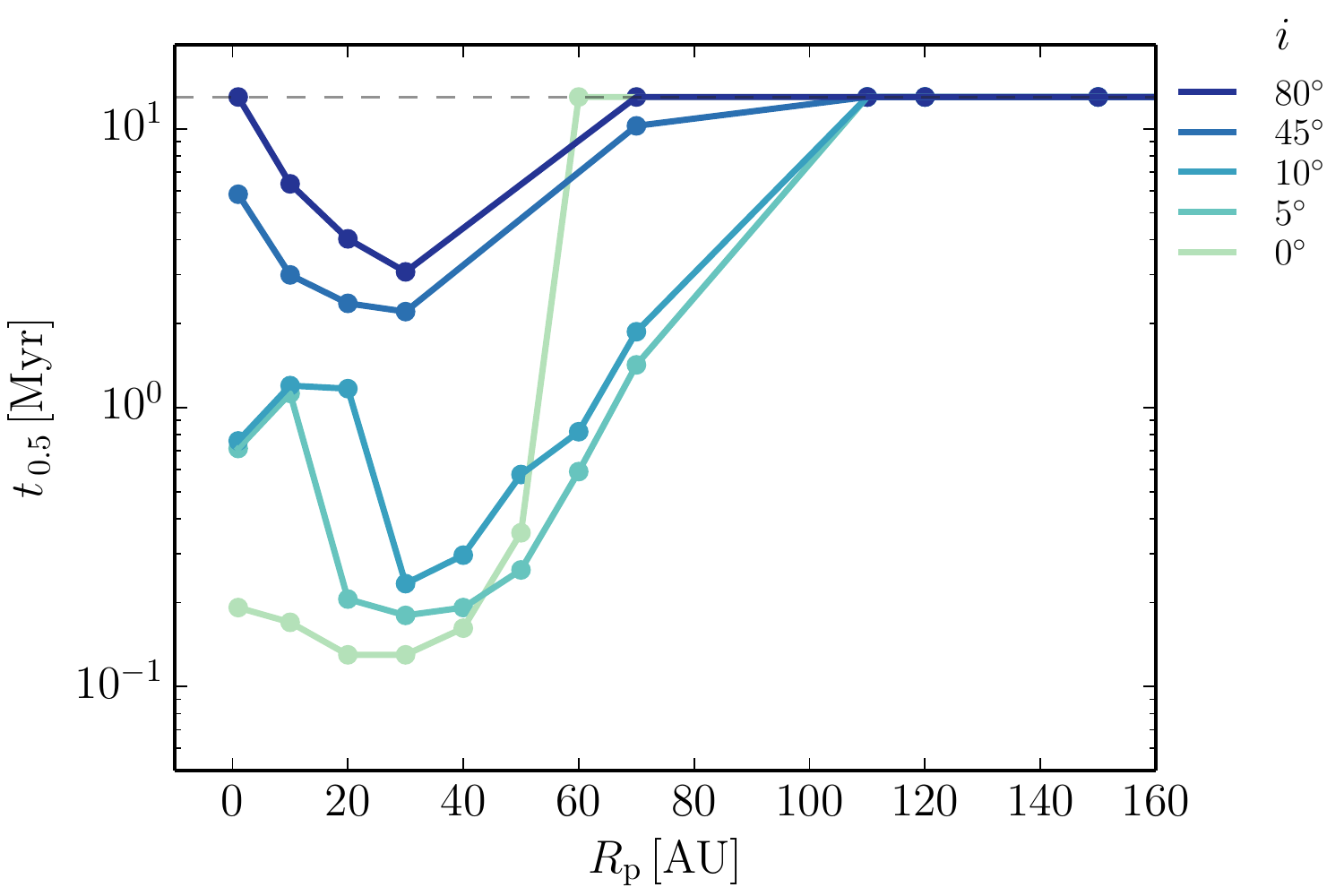}
  \caption{Dependence of $t_{0.5}$, when $f_{\mathrm{d}/\mathrm{b}}(t_{0.5})=0.5$, on the pericenter of the planetary orbit for different inclinations. The dashed horizontal line indicates the lifetime of the system, 13\,Myr.
  }
  \label{fig:t05}
\end{figure}

\begin{figure*}
  \includegraphics[width=0.65\textwidth]{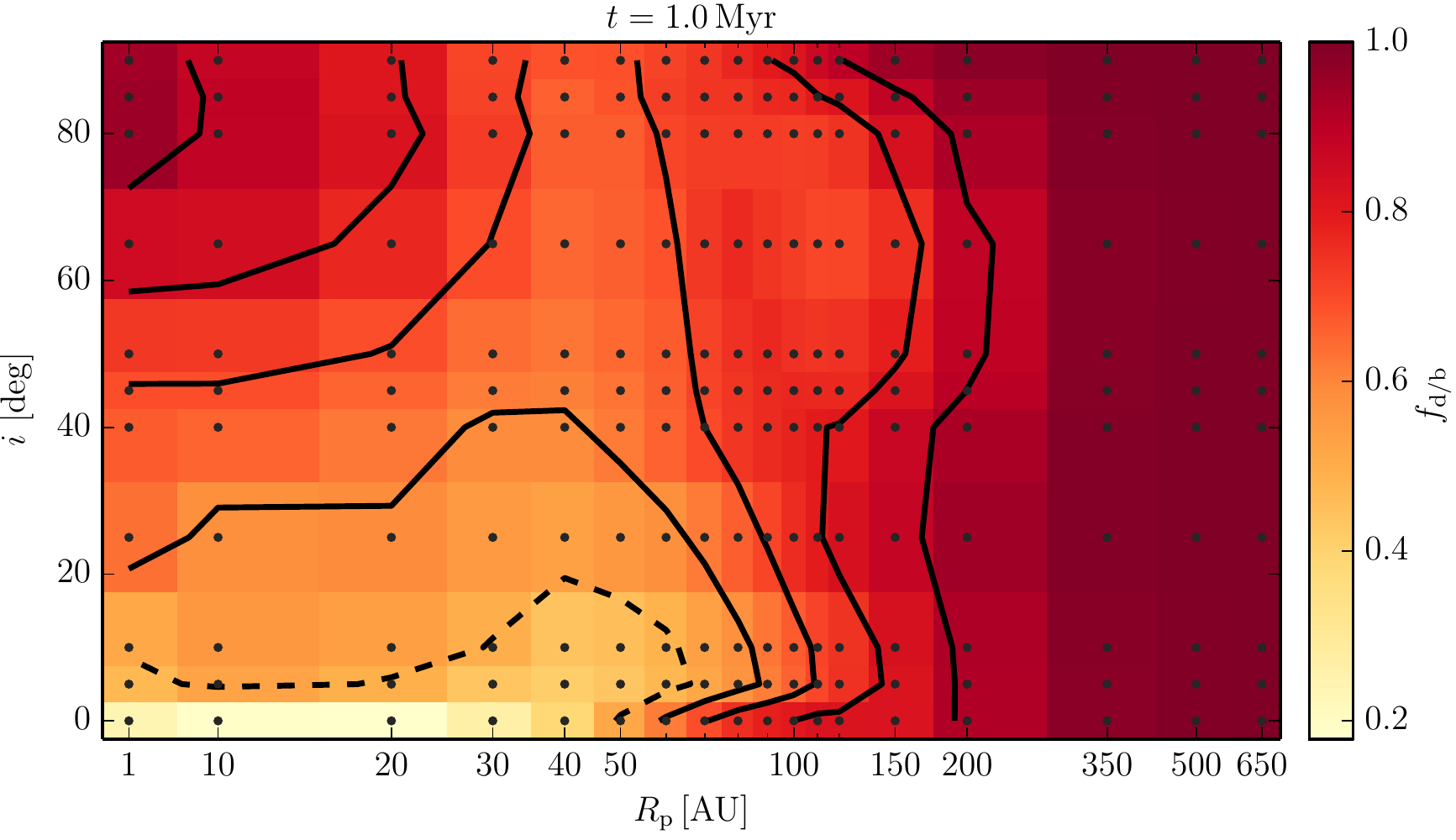}
  \caption{
  The ratio $f_{\mathrm{d}/\mathrm{b}}$ at the time of $1$\,Myr mapped in the pericenter--inclination plane.
  The planetary pericenters and disk inclinations are changing along the horizontal and the vertical axis, respectively.
  The plane is divided in colored bins and the $R_{\rmn{p}}$ and $i$ of the used grid are indicated by points.
  The color maps the $f_{\mathrm{d}/\mathrm{b}}(1\,\rm{Myr})$ for given configuration of $R_{\rmn{p}}$ and $i$. The horizontal axis is logarithmic except for the smallest pericenter (1\,AU), which is shown in a different scale for clarity. Contour lines are over-plotted, their levels go from 0.5 and are increasing by 0.1; the contour for $f_{\mathrm{d}/\mathrm{b}}(1\,\rm{Myr})=0.5$ is indicated by the dashed line.
  }
  \label{fig:ndot_times}
\end{figure*}

When the ratio $f_{\mathrm{d}/\mathrm{b}}$ decreases below 0.5, more bound disk particles are located outside than inside the distance range constrained from observations. 
The moment when $f_{\mathrm{d}/\mathrm{b}}(t_{0.5})=0.5$ can be taken as a measure of the lifetime of the disk as we observe it today.
In Fig.~\ref{fig:t05} we show how $t_{0.5}$ changes with pericenter $R_{\rmn{p}}$ for different inclinations.
Note that for some of the simulations to obtain $t_{0.5}$ for pericenters $R_{\rmn{p}}=1$ and $10$\,AU, $10^3$ particles were used rather than standard $10^4$.
We tested that this does not change the results (see also Sec.~\ref{sec:num_setup}).
In some configurations, $f_{\mathrm{d}/\mathrm{b}}(t)$ is not monotonic and the moment when $f_{\mathrm{d}/\mathrm{b}}=0.5$ occurs more than once (see Sec.~\ref{sec:discussion} for discussion on the oscillations and wobbles) and we use the earliest moment to measure $t_{0.5}$ in these cases.
Using the later times leads to qualitatively similar plot and does not change the conclusions.
Fig.~\ref{fig:t05} shows the $t_{0.5}$ for pericenters up to 150\,AU; wider pericenters, regardless the inclination, have $t_{0.5}$ longer than the system lifetime.

The timescale $t_{0.5}$ is shorter than 1\,Myr for the configurations with low inclination ($i\aplt 10\degr$) and the pericenters smaller and close to the inner edge of the disk ($R_{\rmn{p}}\aplt 60\,$AU). 

The choice of $f_{\mathrm{d}/\mathrm{b}}=0.5$ as the critical value to test for consistency with the observations is arbitrary. 
The appropriate choice is in principle given by the observational limits (i.e., the minimal detectable mass-density of the debris disk).
We verified that the general results do not change when considering a $f_{\mathrm{d}/\mathrm{b}}$ of 0.3--0.8.
As expected, the lower the ratio (i.e., the smaller the fraction of the particles within the original disk region) the longer the timescale.

Values of $f_{\mathrm{d}/\mathrm{b}}$ at $1$\,Myr are shown in Fig.~\ref{fig:ndot_times}. 
Similarly as in Fig.~\ref{fig:t05}, more than half of the bound particles are located outside the disk (i.e., $f_{\mathrm{d}/\mathrm{b}}(1\,\rm{Myr})<0.5$) for the small pericenters and the low inclinations.
The disk stays relatively unperturbed for $R_{\rmn{p}}\apgt 150\,$AU regardless the inclination.

\section{Discussion}
\label{sec:discussion}

\subsection{Disk wobbling and Kozai--Lidov-like oscillations}

\begin{figure*}
  \includegraphics[width=\textwidth]{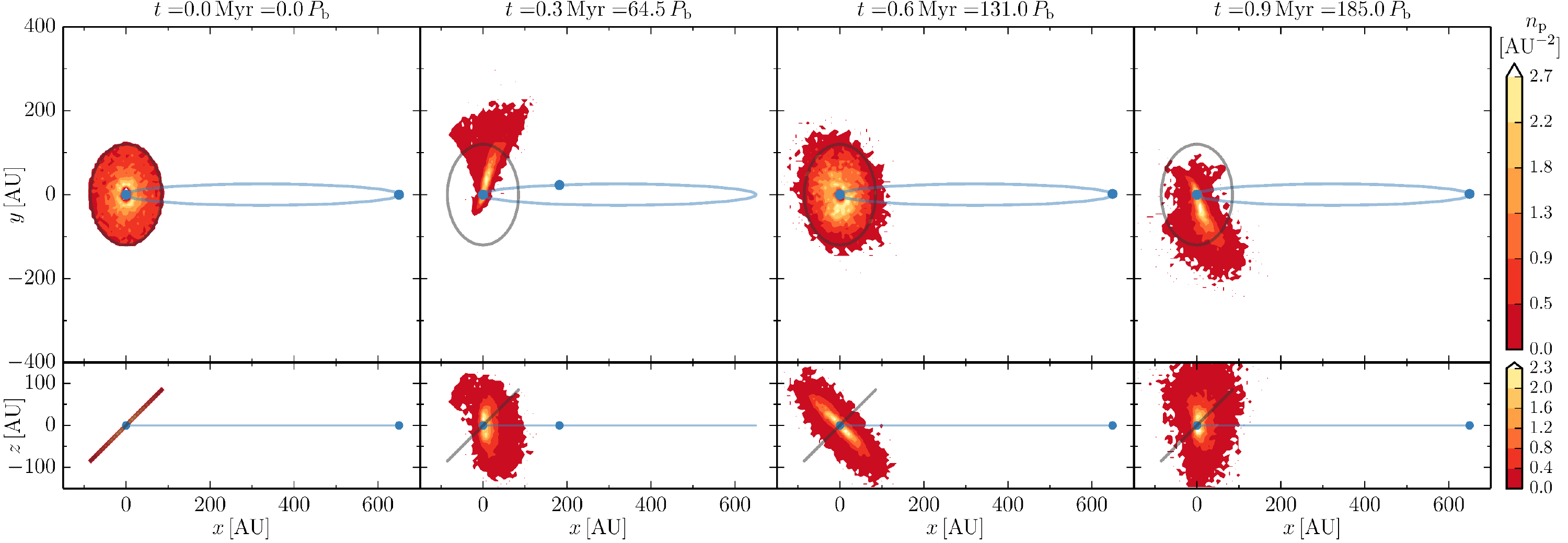}
  \caption{
  Snapshots from the simulation with $R_{\rmn{p}}=1$\,AU and $i=45\degr$; see Fig.~\ref{fig:snaps} for detailed description.
  }
  \label{fig:snaps_1au}
\end{figure*}

As mentioned in Sec.~\ref{sec:disk_lifetime}, for some of the configurations with inclined disks, 
the disk fraction does not decrease monotonously (see Fig.~\ref{fig:ndot_varchar}).
The modulation in $f_{\mathrm{d}/\mathrm{b}}(t)$ can be explained by a wobbling of the disk.
We argue that this wobbling is caused by a mechanism similar to Kozai--Lidov oscillations \citep{kozai_secular_1962,lidov_evolution_1962}.

The Kozai--Lidov mechanism describes exchange of angular momentum in stable hierarchical three-body systems. 
The inner binary is periodically excited to high eccentricity and inclinations with respect to the initial orbital plane, and its argument of periapse librates (i.e., oscillates around a fixed value) with the same period.
However, the energy, i.e., the semi-major axis of the orbit, does not change in the standard picture of the Kozai--Lidov mechanism \citep[e.g.,][]{mardling_tidal_2001}.
The amplitude of the oscillations depends on the relative inclination of the orbits\,---\,the higher the inclination the bigger the changes of eccentricity \citep[e.g.,][]{innanen_kozai_1997}.
The period of the Kozai--Lidov oscillations depends on the masses of the bodies, the periods of the orbits, and the eccentricity of the outer binary. 

The Kozai--Lidov timescale for the restricted three-body problem is approximately 
given by \citep[see, e.g.,][and references therein]{hamers_population_2013},
\begin{equation}
\label{eq:t_kl}
T_{\rmn{KL}} = \alpha\frac{P_{\rmn{b}}^2}{P_{\rmn{d}}} \frac{M_{\star} + M_{\rmn{b}}}{M_{\rmn{b}}} 
\left( 1-e_{\rmn{b}}^2 \right)^{3/2},
\end{equation}
where $P_{\rmn{b}}$ and $e_{\rmn{b}}$ are the period and eccentricity
of the planetary orbit, respectively.  $M_{\star}$ and $M_{\rmn{b}}$
are the central star and the planet mass, respectively.  The orbital period of the disk planetesimal
is $P_{\rmn{d}}$. 
$\alpha$ is a coefficient of order unity.

The strongest modulation of $f_{\mathrm{d}/\mathrm{b}}(t)$ in Fig.~\ref{fig:ndot_varchar} happens for the case with $R_{\rm{p}}=1\,$AU and $i=45^{\circ}$.
This configuration (nor the others presented in Figs.~\ref{fig:ndot_varchar}) does not correspond to the classical Kozai--Lidov example\,---\,the planet orbits inside the inner disk radius and the system star--planet--disk particle does not classify as hierarchical triple.
However, since the planetary orbit is very eccentric (eccentricity of 0.997 for $R_{\rm{p}}=1\,$AU), the time the planet spends closer to the star then 20\,AU, is extremely short\,---\,less than 0.3\% of the orbital period\,---\,and the time within the outer disk radius of 120\,AU is about 3.6\% of the period.
The planet moves outside the disk for most of the time and periodically perturbs the orbits of the disk particles, changing their inclination and eccentricity similarly to the Kozai--Lidov mechanism.
At the same time, we do not observe substantial change in the semi-major axes of planetesimals' orbits and the modulations of $f_{\mathrm{d}/\mathrm{b}}(t)$ are not present when the semi-major axis is used to measure the disk fraction instead the instantaneous distance of the particles from the star.

\begin{figure}
  \includegraphics[width=0.4333\textwidth]{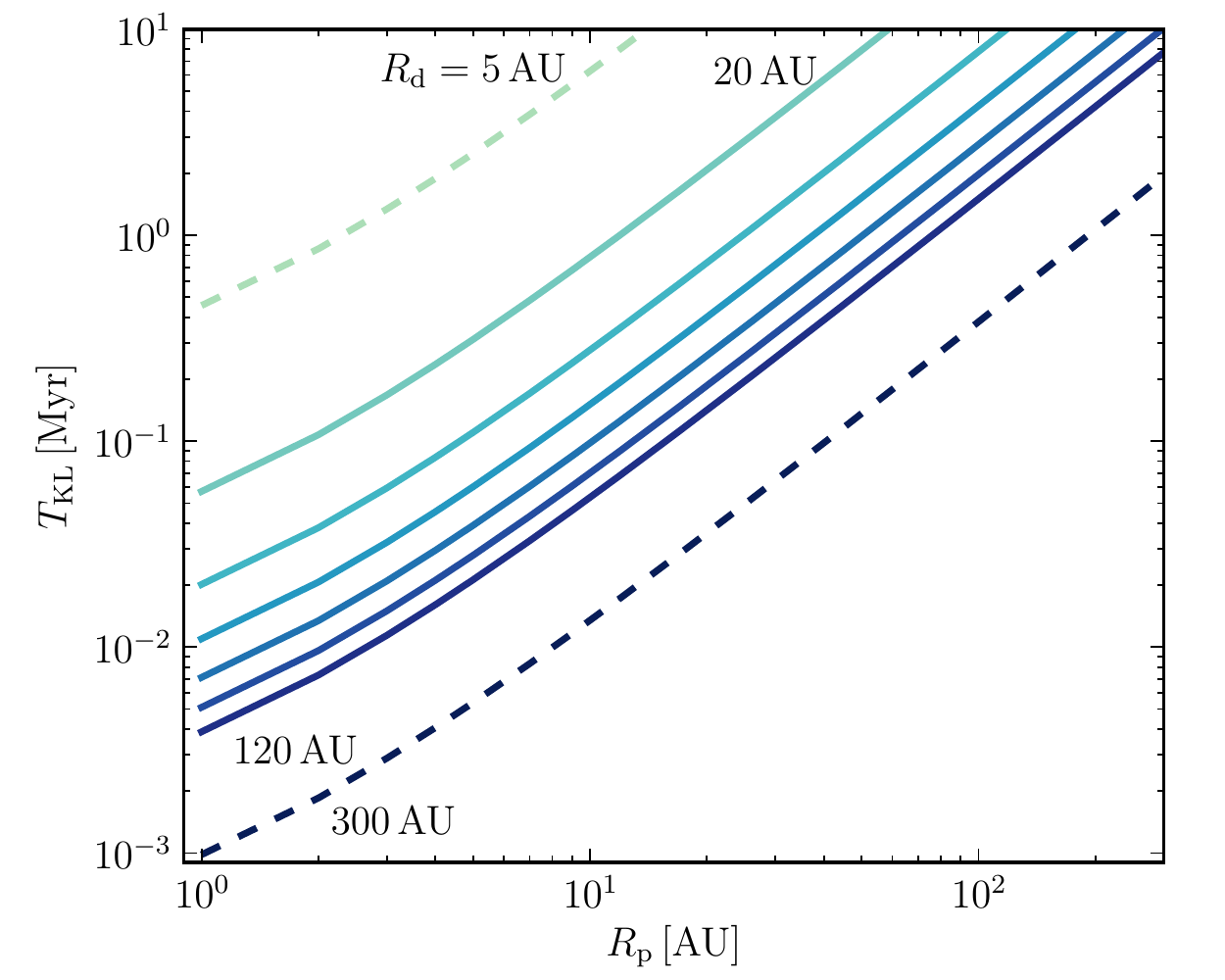}
  \caption{
  Timescale of the Kozai--Lidov mechanism, $T_{\rmn{KL}}$ as given by Eq.~(\ref{eq:t_kl}), as a function of the pericenter of the planetary orbit, $R_{\rmn{p}}$. Different lines show the dependence for different semi-major axes of the disk planetesimals, $R_{\rmn{d}}$. Several values of $R_{\rmn{d}}$ are indicated in the plot. The dashed lines show the cases when $R_{\rmn{d}}$ is outside the initial disk, while the full lines show the cases within the initial disk with a step-size of 20\,AU.
  }
  \label{fig:kozai_per}
\end{figure}

In Fig.~\ref{fig:kozai_per} we show the dependence of $T_{\rmn{KL}}$ on the pericenter of the planetary orbit $R_{\rmn{p}}$ (i.e., on $e_{\rmn{b}}$ and $P_{\rmn{b}}$) for different semi-major axes of the planetesimals $R_{\rmn{d}}$ (i.e., different $P_{\rmn{d}}$).
$T_{\rmn{KL}}$ for $R_{\rm{p}}$ between 1\,AU and 20\,AU ranges from about 0.004 to 1\,Myr depending on $R_{\rmn{d}}$. 
The wobbles happen on the timescale of $\sim 0.1$\,Myr which is generally consistent (considering the factor $\alpha$) with the $T_{\rmn{KL}}$ for the particles in the inner parts of the disk and pericenter $R_{\rm{p}}\sim 1$--5\,AU and the full radial range of the disk for larger $R_{\rm{p}}$.

We suggest that the combination of the perturbation of the planetesimals orbits and a mechanism similar to the Kozai--Lidov oscillations leads to wobbling of the disk, when the eccentricities, inclinations, and the argument of periapse (i.e., the orientation of the orbits) change for a number of disk planetesimals.
We illustrate the process in Fig.~\ref{fig:snaps_1au} where we show snapshots of the simulation with the planetary pericenter at 1\,AU and the disk inclination of $45\degr$.
The four snapshots show the initial state of the system, the times close to the minima ($t=0.3$ and 0.9\,Myr) and maximum ($t=0.6$\,Myr) of the $f_{\mathrm{d}/\mathrm{b}}(t)$ modulation (see Fig.~\ref{fig:ndot_varchar}).
At $t=0.3$ and 0.9\,Myr, the particles are collectively perturbed to higher inclinations and eccentricities and the plane of the disk is close to perpendicular to the orbital plane of the planet, while at $t=0.6$\,Myr, the disk has similar configuration as in the beginning but with retro-grade rotation (inclination of about $-45\degr$).

\subsection{Short-term oscillations of $f_{\mathrm{d}/\mathrm{b}}$}

Apart from the modulation on the timescales of $\sim0.1$\,Myr, 
the disk fractions $f_{\mathrm{d}/\mathrm{b}}(t)$ show periodical modulation with amplitudes of $\aplt 0.03$ and timescales of $\aplt 0.05$\,Myr for most of the configurations (see Figs.~\ref{fig:ndot_varchar} and \ref{fig:ndot_fixed_i}, especially the cases with higher disk fractions).
The modulation results from resonant spiral density waves and rings induced by the planet in the disk.
If a resonant radius is located close to the initial outer edge of the disk, certain number of planetesimals orbit periodically just inside or outside the disk.
The modulation is most prominent for the cases when the relative mass is $f_{\mathrm{d}/\mathrm{b}} \apgt 0.7$ and the resonant patterns are stable enough.
If such resonant features are resolved by future observations, they can provide constraints on the orbit of the planet.

\section{Conclusions}

We studied the lifetime of the debris disk in the peculiar system HD\,106906.  
This 13\,Myr old star is orbited by a debris
disk and a planetary mass companion at a separation of 650\,AU.
We carried out simulations of the system using the AMUSE environment.
Since the disk is much less massive than the star or the planet, we
represent its planetesimals by zero-mass particles.  We implemented a
hybrid numerical method in which the orbit of the planet is solved
independently of the disk and the disk planetesimals are integrated in
the potential of the star and the planet.  The initial conditions for
the simulations were given by the observed characteristics of the
system and the unconstrained characteristics of the system\,---\,namely
the pericenter distance of the planetary orbit and the inclination of
the disk with respect to the planetary orbit\,---\,were systematically
varied.

We find that more than 80\% of the disk particles stay bound to the star for majority of the considered configurations and only in the case of orbits with low inclination $\aplt 10^{\circ}$ and pericenter of the planetary orbit $\aplt50\,$AU, a substantial part of the disk is lost during the first 1\,Myr of the evolution.
To estimate how long the disk stays in a configuration consistent with the observations, we followed the ratio of the number of the disk particles with distance within the constrained disk radii (20--120\,AU) and the number of the particles bound to the system. 
We define the lifetime of the disk when more particles are orbiting outside than within the constrained disk radius (i.e., more particles have is at distance $<20$\,AU or $>120$\,AU from the star). 
The lifetime of the disk is shorter than 1\,Myr for orbits with low inclination $i<5^{\circ}$ and comparable with 1\,Myr when $i\sim 5$--$10^{\circ}$, and with pericenter smaller or close to the inner edge of the disk ($R_{\rm{p}} \aplt 50\,$AU, see Figs.~\ref{fig:t05} and~\ref{fig:ndot_times}).
Such orbits are expected in the case when the planet formed closer to the star, most probably within the inner disk edge where it cleared the inner region, and was scattered to its current orbit by other member of the system.
However, such interaction is estimated to occur during the first 10\,Myr of the lifetime of planetary systems \citep[e.g.,][]{veras_formation_2009,scharf_long-period_2009}. 
Considering the current age of the system of $13\pm2$\,Myr \citep{pecaut_revised_2012}, we conclude that the configurations with lifetimes shorter than 1\,Myr ($i \aplt 10^{\circ}$ and $R_{\rm{p}} \aplt 50\,$AU) are inconsistent with the scenario according to which the current orbit resulted from planet--planet scattering from the inner disk.

When the disk is inclined with respect to the planetary orbit with inclination $\apgt40\degr$, it can survive longer than 1\,Myr even in case the pericenter is within the inner disk edge.
In these configurations, the disk wobbles (see Fig.~\ref{fig:snaps_1au}).
We argue that this is caused by a mechanism similar to the Kozai--Lidov oscillations induced by the
planet on the disk particles.
The planet can also induce resonant structures in the disk, such as spiral arms and rings.

\section*{Acknowledgments}

We thank the anonymous referee for reviewing our work and for insightful comments which improved the manuscript.
We thank Guilherme Gon\c{c}alves Ferrari, Inti Pelupessy, and Tjarda Boekholt for their help with the Kepler solver used in our numerical method.  
We thank Adrian Hamers for enriching discussions about the Kozai--Lidov mechanism, and Matthew Kenworthy and Tiffany Meshkat for discussions about HD\,106906. 
We thank Gr\'{a}inne Costigan for reading the manuscript and for her useful comments.
This work was supported by the Interuniversity Attraction Poles Programme initiated by the Belgian
Science Policy Office (IAP P7/08 CHARM) and by the Netherlands Research Council NWO (grants \#643.200.503, \#639.073.803 and \#614.061.608) and by the Netherlands Research School for Astronomy (NOVA).

\bibliographystyle{mn2e}
\bibliography{spdd}

\label{lastpage}
\end{document}